\documentclass[aps,pra,twocolumn,groupedaddress,showpacs]{revtex4}

\usepackage[latin1]{inputenc}
\usepackage[T1]{fontenc}
\usepackage{amsmath,amssymb}
\usepackage{bbm}
\usepackage{graphicx}
\usepackage{subfigure}
\usepackage{sistyle}
\usepackage{ifpdf}

\DeclareMathOperator{\diag}{diag}
\DeclareMathOperator{\var}{Var}

\newcommand\relphantom[1]{\mathrel{\phantom{#1}}}

\begin{document}

\title{Estimation of fluctuating magnetic fields by an atomic magnetometer}

\author{Vivi Petersen and Klaus M\o lmer}
\affiliation{QUANTOP - Danish National Research Foundation Center for
  Quantum Optics\\
  Department of Physics and Astronomy, University of
  Aarhus, DK-8000 Århus C, Denmark}

\date{\today}

\begin{abstract}
  We present a theoretical analysis of the ability of atomic
  magnetometers to estimate a fluctuating magnetic field. Our analysis
  makes use of a Gaussian state description of the atoms and the
  probing field, and it presents the estimator of the field and a
  measure of its uncertainty which coincides in the appropriate limit
  with the achievements for a static field. We show by simulations
  that the estimator for the current value of the field systematically
  lags behind the actual value of the field, and we suggest a more
  complete theory, where measurement results at any time are used to
  update and improve both the estimate of the current value and the
  estimate of past values of the B-field.
\end{abstract}


\maketitle

\section{Introduction}
\label{sec:introduction}

For decades superconducting quantum interference devices, SQUIDs, have
been unrivalled as the most sensitive detectors of weak magnetic
fields, and they have been applied in diverse scientific studies
including NMR signal detection~\cite{greenberg98:_applic},
visualization of human brain activity~\cite{rodriguez99}, and
gravitational wave detection~\cite{harry00:_two_q}. Optically pumped
atomic gasses offer an alternative means to detect weak magnetic
fields via the induced Larmor precession of the polarized spin
component, and the possibility to avoid the need for cryogenic cooling
and the relatively high price of the SQUID devices, has spurred an
interest in employing atomic magnetometers in medical diagnostics, see
for example work on the mapping of human cardiomagnetic
fields,~\cite{bison03:_dynam}. Combined with the fact, that the atomic
based magnetometers may now reach superior field
sensitivity~\cite{kominis03}, we are now highly motivated to
investigate and identify the optimal performance and fundamental
limits of such devices.

Atoms constitute ideal probes for a number of physical phenomena, and
since they are quantum systems, the statistical analysis of
measurement results has to take into account the very special role of
measurements in quantum theory. In high precision metrology the aim is
to reduce error bars as much as possible, and there is an obvious
interest in making the tightest possible conclusion from measurement
data. How to optimally prepare and interact with a physical system to
obtain maximum information, and even the simpler task of identifying
precisely the information available from a specific (noisy) detection
record are not fully characterized at this moment. Large efforts are
currently being made to combine techniques from classical control and
parameter estimation theory with quantum filtering equations.

In the present paper we analyze the case of atomic magnetometry with
large atomic samples which are being probed continuously in time with
weak optical probes. Such systems allow a simplified treatment by
means of Gaussian states and probability distributions, and we are
able to give exact expressions for the probability distribution of the
probed magnetic field, i.e., our result is neither too weak nor too
strong (no procedure exists by which further information can be
extracted from the available data, and the actual value must agree
with our estimate within the probability distribution). Our work is a
generalization of previous work for static
fields~\cite{molmer04:_estim_param_gauss_probes,
  geramia03:_quant_kalman_filter_heisen_limit_atomic_magnet} to the
interesting case of time dependent fields. Time dependent fields were
studied recently~\cite{stockton04:_robus_quant_param_estim}, and we
shall generalize that analysis and show that further improvements of
the estimate of the field at a given time is possible by taking into
account the detailed detection record at both earlier and later times.

In Sec.~\ref{sec:how-estimate-b}, we describe the physical setup and
the interactions between a single component $B$-field , the atomic
system and the optical probe field. In Sec.~\ref{sec:devel-atoms-due},
we describe the time evolution of the collective atomic quantum state
during interaction with a given, parametrized, field $B(t)$ and
optical probing. In Sec.~\ref{sec:measure-b-field}, we treat the case,
where the time dependent magnetic field is assumed to be a realization
of an Ornstein-Uhlenbeck stochastic process, where the actual time
dependence of $B(t)$ is unknown by the experimentalist, and we provide
the formalism for the optimum estimate for $B(t)$ based on the noisy
detection record for all earlier times $t'<t$. In
Sec.~\ref{sec:results}, we show results of numerical simulations,
suggesting that also the detection at later times $t'>t$ improves our
estimate of $B(t)$, and we present \textit{ad hoc} and precise
analyses of the optimum estimator. Sec.~\ref{sec:discussion} concludes
the paper.

\section{Interaction between a large atomic sample, a magnetic field
  and an optical probe}
\label{sec:how-estimate-b}

\subsection{Physical System and Interactions}
\label{sec:phys-syst-inter}

We consider a gas with a macroscopic number $N_\mathrm{at}$ of atoms
with two degenerate Zeeman states. Initially, all the atoms are
prepared by optical pumping in the same internal quantum state, and
all interactions are assumed to be invariant under permutations of the
atoms. The dynamics is conveniently described by the collective
effective spin operator $\mathbf{J} = \frac{\hbar}{2}\sum_i
\boldsymbol{\sigma}_i$ with $\boldsymbol{\sigma}_i$ the Pauli matrices
describing the individual two-level atoms. The atoms are initially
prepared by optical pumping such that their spin is polarized along
the $x$ axis, and we assume only a small depolarization during the
interaction, so that the operator $J_x$ can be well approximated by a
constant number $\langle J_x\rangle = \frac{\hbar N_\mathrm{at}}{2}$.
The other two projections of the collective spin, $J_y$ and $J_z$,
obey the commutation relation $[ J_y, J_z] = i \hbar J_x$ and the
resulting uncertainty relation on $J_y$ and $J_x$, i.e., on the number
of atoms populating the $\sigma^y$ and $\sigma^z$ atomic eigenstates
precisely reflect the binomial distribution of atoms on these states.
The commutator may be rewritten as $[x_\text{at}, p_\text{at}] = i$
for the effective canonical position and momentum variables $
x_\mathrm{at} = \frac{J_y}{\sqrt{\hbar\langle J_x\rangle}},
p_\mathrm{at} = \frac{J_z}{\sqrt{\hbar\langle J_x\rangle}}$, and the
binomial population statistics of the collective states nicely maps to
Gaussian probability distributions of $x_\text{at}$ and $p_\text{at}$.

\begin{figure}
  \centering
  \includegraphics{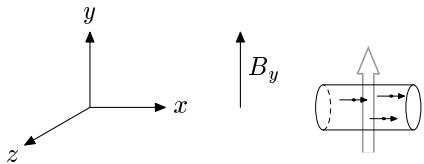}
  \caption{A $B$-field component along the $y$-axis causes a Larmor
    precession of an atomic spin which is initially polarized along
    the $x$-axis. The resulting $z$ component of the spin is probed by
    the Faraday polarization rotation of a linearly polarized laser
    beam which propagates along the $y$-axis.}
  \label{fig:1}
\end{figure}

We now imagine that the atomic sample is placed in a $B$-field
directed along the $y$ direction, which causes a Larmor rotation of
the atomic spin towards the $z$ axis. The setup is shown in
Fig.~\ref{fig:1}. During a time interval $\tau$,
the $z$-component of the collective spin, represented by the operator
$p_\text{at}$, evolves as
\begin{gather}
  \label{eq:36}
  p_\text{at} \to p_\text{at} - \mu\tau B,
\end{gather}
where $\mu$ is given by the magnetic moment $\beta$, via $\mu =
(1/\hbar)\beta\sqrt{\langle J_x\rangle/\hbar}$. By probing the value
of $p_\text{at}$, we acquire information about the magnetic field.

A number of papers have dealt with the interaction of atomic samples
with an optical
field,~\cite{kuzmich98:_atomic_quant_non_demol_measur_squeez,
  takahashi99:_quant_farad,
  kuzmich99:_quant_nondem_measur_collec_atomic_spin,
  duan00:_quant_commun_atomic_ensem_using_coher_light,
  julsgaard01:_exper_long_entan_two_macros_objec,
  molmer04:_estim_param_gauss_probes,
  madsen04:_spin_squeez_precis_probin_light2}. A particularly
convenient interaction occurs when the two atomic states interact with
a linearly polarized non-resonant probe field. A linearly polarized
field can be expanded on two circular components, and if they interact
differently with the two internal atomic states, they experience
different phase shifts depending on the atomic populations, which in
turn translates into a (Faraday) rotation of the polarization of the
field. For use of the effective spin-$1/2$ picture to the description
of angular momenta larger than $1/2$, see for
example~\cite{sherson06:_deter}. We are interested in the case of a
continuous beam of light interacting with the atoms, and this is
effectively obtained by discretizing the field into a sequence of
square pulse segments of light,
see~\cite{molmer04:_estim_param_gauss_probes}%(\textbf{XXX} Lars og
                                %Klaus magnetometer chapter)
. Since the light beam is linearly
polarized along $x$, the Stokes operator for a short segment of the
beam with $N_\mathrm{ph}$ photons has an x-component which behaves
classically, $\langle S_x\rangle = \frac{\hbar N_\mathrm{ph}}{2}$, and
the two remaining components fulfill the commutator relation of
angular momentum operators. Accordingly, for the scaled effective
variables $ x_\mathrm{ph} = \frac{S_y}{\sqrt{\hbar\langle
    S_x\rangle}}, p_\mathrm{ph} = \frac{S_z}{\sqrt{\hbar\langle
    S_x\rangle}}$, we have $[x_\text{ph}, p_\text{ph}] = i$ and the
initial coherent state of the field is a minimum uncertainty Gaussian
state in these variables. When a single light segment is sent through
the atomic sample, the polarization rotation is quantified by the
difference between the intensities of linear polarization components
at $45^\circ$ and $-45^\circ$ angles with respect to the incident
polarization, i.e., by the $S_y$ $(x_\mathrm{ph})$ quantum variable.
The field is not in an eigenstate of this quantity, and the
measurement outcome is stochastic with a mean value governed by the
mean rotation, i.e., the mean atomic population difference, and a
variance given by a shot noise contribution from the photon number
distribution, and by the variance of the atomic populations. In turn,
the detection of the light signal causes a back-action on the quantum
state of the atoms, which, depending on the total number of photons
detected, shifts the mean atomic populations and reduces their widths.

As shown in Ref.\cite{madsen04:_spin_squeez_precis_probin_light2}
and references therein, the polarization rotation described above is
described by a simple effective Hamiltonian, and in the Heisenberg
picture the atomic variables and the variables describing a single
segment of the light beam of duration $\tau$ evolve as
\begin{align}
  \label{eq:evo1}
  x_\text{at} &\mapsto x_\text{at} + \kappa \sqrt{\tau}
  p_\text{ph}, &p_\text{at} \mapsto p_\text{at} \\
  \label{eq:evo2}
  x_\text{ph} &\mapsto \kappa \sqrt{\tau} p_\text{at} +
  x_\text{ph}, & p_\text{ph} \mapsto p_\text{ph}
\end{align}
where the atom-light coupling strength $\kappa =
\frac{d^2\omega}{\Delta Ac\epsilon_0}
\sqrt{N_\mathrm{at}N_\mathrm{ph}/\tau}$ is a function of atomic
parameters ($d$ is the atomic dipole moment, $\omega$ is the photon
frequency, $\Delta$ is the detuning of the light from atomic
resonance, and $A$ is the area of the the light field) and, notably,
the total number of atoms $N_\mathrm{at}$ and the photon flux $\Phi =
N_\mathrm{ph}/\tau$ in the pulse segment. Our theory is to a large
extent analytical, but in order to present some of the results in
graphs, we shall adopt realistic values, summarized in Table 1, for a
possible experiment with cesium atoms.
\begin{table}%[htbp]
  \caption{Values of physical quantities used in the numerical
    examples of this paper.}
  \label{tab:1}
  \begin{ruledtabular}
    \begin{tabular}{ll}
      Physical quantity & Value\\
      \hline
      Initial uncertainty & $\Delta B_0 = \SI{1}{pT}$\\
      Total number of atoms & $N_\mathrm{at} = 2\times10^{12}$\\
      Photon flux & $\Phi = \SI{5e14}{s^{-1}}$\\
      Intersection area & $A = \SI{2}{mm^2}$\\
      Wavelength & $\lambda = \SI{852}{nm}$\\
      Detuning & $\frac{\Delta}{2\pi} = \SI{10}{GHz}$\\
      Atomic dipole moment & $d = \SI{2.61e-29}{Cm}$\\
      Coupling between atoms and $B$-field & $\mu =
      \SI{8.79e4}{s^{-1}}$\\
      Coupling between atoms and photons & $\kappa^2 =
      \SI{1.83e6}{s^{-1}}$\\
    \end{tabular}
  \end{ruledtabular}
\end{table}

\subsection{Gaussian state formalism}
\label{sec:devel-atoms-due}

In this section, we shall assume a magnetic field $B(t)$ with an
explicitly given time dependence. This analysis will be needed, when
we proceed to simulate how an unknown field is estimated, since it is
the actual realization of the field, that drives the atomic dynamics.
The current section thus accounts for the ``information available to
the theorist'', whereas the proceeding section will deal with the
``information available to the experimentalist'' having only access to
the optical detection record.

We place a spin-polarized atomic gas in the $B$-field, and probe it
continuously by the polarized light beam. The Gaussian probability
distribution (Wigner function) of the atomic variables and of the
light pulse prior to interaction, evolves into a new Gaussian state,
and the detection of the polarization rotation, which is a measurement
of the field observable $x_\text{ph}$ leads to an update of the atomic
state, but it retains the Gaussian
form~\cite{eisert03:_introd_basic_entan_theor_contin_variab_system2}.
We can hence describe the atomic state by the mean values and by the
covariance matrix which fully characterize a Gaussian state, and we
shall now summarize the dynamics of the atoms+fields with an effective
formalism that describes the time evolution of the mean values and the
covariances of the atomic state due to interaction and measurements.

We first define a vector of variables (operators) $\mathbf{y} =
(x_\text{at}, p_\text{at}, x_\text{ph}, p_\text{ph})^T$, with the
corresponding vector of mean values $\mathbf{m}=\langle
\mathbf{y}\rangle$ and with the covariance matrix
$\boldsymbol{\gamma}$ where $\gamma_{ij} = 2 \text{Re} \langle
(y_i-\langle y_i\rangle)(y_j-\langle y_j\rangle)\rangle$. Our Gaussian
state description of the spin polarized sample and the linearly
polarized light translates into the specification of the initial
values,
\begin{align}
  \mathbf{m} &= 0\label{eq:23}\\
  \boldsymbol{\gamma} &= \mathbbm{1}_{4\times4}.\label{eq:22}
\end{align}
and the update formula due to interactions during a time interval $\tau$
\begin{align}
  \mathbf{m} &\to \mathbf{Sm} + \mathbf{v}\label{eq:24}\\
  \boldsymbol{\gamma} &\to
  \mathbf{S}\boldsymbol{\gamma}\mathbf{S}^T\label{eq:25}
\end{align}
where
\begin{gather}
  \label{eq:26}
  \mathbf{S} =
  \begin{pmatrix}
    1 & 0 & 0 & \kappa \sqrt{\tau}\\
    0 & 1 & 0 & 0\\
    0 & \kappa \sqrt{\tau} & 1 & 0\\
    0 & 0 & 0 & 1
  \end{pmatrix}
\end{gather}
and
\begin{gather}
  \label{eq:27}
  \mathbf{v} = (0, -\mu\tau B(t), 0, 0)^T.
\end{gather}

Let us write
\begin{align}
  \boldsymbol{\gamma} &=
  \begin{pmatrix}
    \mathbf{A}_\gamma & \mathbf{C}_\gamma\\
    \mathbf{C}_\gamma^T & \mathbf{B}_\gamma
  \end{pmatrix}\label{eq:28}\\
  \mathbf{m} &= (\mathbf{m}_A, \mathbf{m}_B)\label{eq:34}
\end{align}
with $\mathbf{A}_\gamma$ the covariance matrix for the atomic
variables, $\mathbf{y}_1 = (x_\text{at},p_\text{at})^T$,
$\mathbf{B}_\gamma$ the covariance matrix for the field variables,
$\mathbf{y}_2 = (x_\text{ph},p_\text{ph})^T$, and $\mathbf{C}_\gamma$
the correlation matrix between $\mathbf{y}_1$ and $\mathbf{y}_2^T$.
When a measurement of the variable $x_\text{ph}$ is performed, the
outcome takes on a random value, given by the Gaussian probability
distribution. For a short segment of light, the mean value of
$x_{\textrm{ph}}$ is $m_3=\kappa\sqrt{\tau}m_2$, and the variance is
one half (the incident field variance is only infinitesimally modified
by the atoms). The measurement of a value $x_\text{meas}$ for
$x_\text{ph}$ collapses the field state and transforms the atomic
component according to
\begin{align}
  \label{eq:29}
  \mathbf{A}_\gamma &\to \mathbf{A}_\gamma
  - \mathbf{C}_\gamma(\pi\mathbf{B}_\gamma\pi)^-
  \mathbf{C}_\gamma^T\\
  \label{eq:32}
   \mathbf{m}_A &\to \mathbf{m}_A
  + \mathbf{C}_\gamma (\pi\mathbf{B}_\gamma\pi)^-((x_\text{meas}-m_3),0)^T\\
  \label{eq:30}
  \mathbf{B}_\gamma &\to \mathbbm{1}_{2\times2}\\
  \label{eq:31}
  \mathbf{C}_\gamma &\to 0\\
  \label{eq:33}
  \mathbf{m}_B &\to 0
\end{align}
where $\pi=\diag(1,0)$ and $(\dots)^-$ denotes the Moore-Penrose
pseudoinverse. To the lowest, relevant order,
$(\pi\mathbf{B}_\gamma\pi)^- = \textrm{diag}(1,0)$.
Equations~\eqref{eq:30},~\eqref{eq:31}, and~\eqref{eq:33} ``refresh''
the atom and field variables corresponding to the subsequent light
segment which has no correlations with the atoms prior to interaction.
Note that the quantity $(x_\text{meas}-m_3)$ is a random variable with
vanishing mean and variance $1/2$, and since $\mathbf{C}_\gamma$
scales with $\sqrt{\tau}$, the measurement induced, random
displacement of the mean values~\eqref{eq:32} can also be expressed in
terms of a Wiener increment with zero mean and variance $\tau$,
cf.~\cite{geramia03:_quant_kalman_filter_heisen_limit_atomic_magnet,
  stockton04:_robus_quant_param_estim}.

Before proceeding to the interaction between the atoms and an unknown
magnetic field, we note that the dynamics in~\eqref{eq:evo1} is of
quantum non demolition type (QND), i.e., it permits a detection of the
atomic variable $p_\text{at}$ without changing this observable. Such a
detection (carried out when the field variable $x_\text{ph}$ is read
out after the atom-light interaction), in effect leads to squeezing of
the $p_\text{at}$ component of the atomic spin around a random value,
which is selected by the random outcome of the measurement process and
by the deterministic Larmor rotation. Of course a single,
infinitesimal time step as in~(\ref{eq:evo1},\ref{eq:evo2}) leads only
to an infinitesimal squeezing, but as the evolution proceeds
continuously in time it leads to a monotonic reduction of the variance
of the atomic spin component. If atomic spontaneous decay is taken
into account, the squeezing has an optimum and further interaction
with the optical probe leads to incoherent depopulation among the
atomic states. We are indeed able to describe also such processes in
the Gaussian state formalism~\cite{molmer04:_estim_param_gauss_probes,
  madsen04:_spin_squeez_precis_probin_light2}, but since the
importance of spontaneous processes is very system specific, and it
can be reduced by going to sufficiently large detunings, we shall
ignore spontaneous decay in the following.

\section{Estimation of an unknown time dependent $B$-field}
\label{sec:measure-b-field}

The $B$-field causes the Larmor precession~\eqref{eq:evo1}. By probing
the value of $p_\text{at}$, we acquire information about the magnetic
field if its value is not already known.
In~\cite{molmer04:_estim_param_gauss_probes,
  petersen05:_magnet_entan_atomic_sampl}, we found that a constant
magnetic field is effectively probed with a time dependent variance on
the estimate given by
\begin{equation}
  \label{eq:bvaranalyt}
  \Delta B(t)^2 = \frac{\Delta B_0^2 (1+ \kappa^2 t)} {1+\kappa^2 t +
    \frac{2}{3} \kappa^2 \mu^2 (\Delta B_0)^2 t^3 + \frac{1}{6} \kappa^4
    \mu^2 (\Delta B_0)^2 t^4}
\end{equation}
with $\Delta B_0^2$ representing our prior knowledge of the field. In
the limit of $\kappa^2 t \gg 1$, we have $\Delta B(t)^2 \simeq
6/(\kappa^2\mu^2 t^3)$ which is independent of the prior knowledge and
which reflects a more rapid reduction with time of the variance than
expected from a conventional statistical argument. This is due to the
atomic squeezing, as it progressively makes the system more and more
sensitive to the magnetic field perturbation.

In the following, we shall generalize the analysis to the case of time
dependent $B$-fields. A convenient model for a random field is a
damped diffusion (Ornstein-Uhlenbeck) process, governed by the
stochastic differential equation
\begin{gather}
  \label{eq:1}
  dB(t) = -\gamma_b B(t)dt + \sqrt{\sigma_{b}}dW_b
\end{gather}
where the Wiener increment $dW_b$ has a Gaussian distribution with
mean zero and variance $dt$. Our task is to expose atoms to a
realization of this process and to use the polarization measurements
to construct an estimate for the actual current value of the field.

The Ornstein-Uhlenbeck process can be simulated on a computer, but we
can also make statistical predictions, e.g., the steady state mean
vanishes and the variance is
\begin{equation}
  \label{eq:steadyou}
  \textrm{Var}_{\textrm{st}}(B)=\frac{\sigma_{b}}{2\gamma_b},
\end{equation}
and if the value of the field is estimated at time $t_L$ in the
laboratory to be $\textrm{B}_{t_L}$ with a variance $V_{t_L}$ on the
estimate, our best estimate for the value at a \textit{future} time
$t>t_L$ takes the value $\textrm{B}_t \exp(-\gamma_b(t-t_L))$ with a
variance
\begin{equation}
  \label{eq:futureou}
  V_{t} = V_{t_L} e^{-2\gamma_b(t-t_L)}
  + \frac{\sigma_b}{2\gamma_b}(1-e^{-2\gamma_b(t-t_L)}).
\end{equation}

The Ornstein-Uhlenbeck process can be very slow, in which case the
field retains its random value almost constantly over long times, and
we expect to recover the results
in~\cite{molmer04:_estim_param_gauss_probes} for estimation of a
constant field, because the accumulated photo detection record over
time carries information about the time dependent field, which is
known to differ not very much. If the Ornstein-Uhlenbeck process is
very fast, the accumulated photo detection record until the present
time $t$ gives only little information about the present value. Note
that by scaling $\gamma_b$ and $\sigma_b$ by the same factor, the
variance~\eqref{eq:steadyou} is unchanged, but the process changes from
slow to rapid fluctuations.

Our theoretical description of the estimation process deals with a
joint Gaussian distribution for the quantum variables and the
classical magnetic field. As
in~\cite{molmer04:_estim_param_gauss_probes} we formally treat the
B-field as the first component in our vector of five Gaussian
variables $\mathbf{\widetilde{y}} = (B, x_\text{at}, p_\text{at},
x_\text{ph}, p_\text{ph})$, where the tilde is used to distinguish
these variables from those in the previous section. The Gaussian state
is characterized by its mean value vector
$\mathbf{\widetilde{m}}=\langle \widetilde{y}\rangle$ and its
covariance matrix $\boldsymbol{\widetilde{\gamma}}$ where
$\widetilde{\gamma}_{ij} = 2 \text{Re} \langle
(\widetilde{y}_i-\langle
\widetilde{y}_i\rangle)(\widetilde{y}_j-\langle
\widetilde{y}_j\rangle)\rangle$.

Note that although, e.g., $x_\text{at}$ is the same operator in this
and the previous section, its Gaussian state mean value and variance
are not the same, because the Gaussian state is the probability
distribution assigned by the observer given his or her acquired
knowledge about the system. In the previous section we determined this
probability distribution conditioned on full knowledge of the time
dependent $B$-field and the photo detection record, whereas in the
current section, only ``the experimentalist's'' knowledge of the
detection record is assumed.

By propagating the update formulas for $\mathbf{\widetilde{m}}$ and
$\boldsymbol{\widetilde{\gamma}}$ due to the Larmor rotation, the
atom-light interaction, the random outcomes of the probing process,
and the Ornstein-Uhlenbeck process, we obtain at any time a mean
value $\widetilde{m}_1$ and a variance $\widetilde{\gamma}_{11}/2$
for the $B$-field.

The initial values are
\begin{align}
  \mathbf{\widetilde{m}} &= 0\label{eq:3}\\
  \boldsymbol{\gamma}_0 &= \diag(2\var(B_0),1,1,1,1).\label{eq:4}
\end{align}
Due to the interaction between the $B$-field and the atoms and
between the atoms and the photons we have
\begin{align}
%  \mathbf{\widetilde{y}} &\to \mathbf{S}_1\mathbf{\widetilde{y}}\label{eq:10}\\
  \mathbf{\widetilde{m}} &\to \mathbf{S}_1\mathbf{\widetilde{m}}\label{eq:11}\\
  \boldsymbol{\widetilde{\gamma}} &\to
  \mathbf{S}_1\boldsymbol{\widetilde{\gamma}}\mathbf{S}_1^T\label{eq:12}
\end{align}
where
\begin{gather}
  \label{eq:9}
  \mathbf{S}_1 =
  \begin{pmatrix}
    1 & 0 & 0 & 0 & 0\\
    0 & 1 & 0 & 0 & \kappa \sqrt{\tau}\\
    -\mu \tau & 0 & 1 & 0 & 0\\
    0 & 0 & \kappa \sqrt{\tau} & 1 & 0\\
    0 & 0 & 0 & 0 & 1
  \end{pmatrix}
\end{gather}
In this section the $B$ field is one of the variables, and hence the
equation~\eqref{eq:11} is linear unlike the affine
transformation~\eqref{eq:24}. Due to the Ornstein-Uhlenbeck process we
have
\begin{align}
  \widetilde{m}_1 &\to \widetilde{m}_1(1-\gamma_b\tau)\label{eq:6}\\
  \boldsymbol{\widetilde{\gamma}} &\to
  \mathbf{S}_2 \boldsymbol{\widetilde{\gamma}} \mathbf{S}_2
  + \mathbf{L}\label{eq:8}
\end{align}
where $\mathbf{S}_2 = \diag(1-\gamma_b\tau,0,0,0,0)$ and $\mathbf{L} =
\diag(\sigma_{b}\tau,0,0,0,0)$.

To handle the measurement on $\widetilde{x}_\text{ph}$ we write the
covariance matrix and mean value vector as
\begin{align}
  \boldsymbol{\widetilde{\gamma}} &=
  \begin{pmatrix}
    \mathbf{\widetilde{A}}_\gamma & \mathbf{\widetilde{C}}_\gamma\\
    \mathbf{\widetilde{C}}_\gamma^T & \mathbf{\widetilde{B}}_\gamma
  \end{pmatrix}\label{eq:13}\\
  \mathbf{m} &= (\mathbf{m}_A, \mathbf{m}_B)\label{eq:35}
\end{align}
with $\mathbf{\widetilde{A}}_\gamma$ the covariance matrix for the
$B$-field and atoms, $\mathbf{\widetilde{y}}_1 =
(B,\widetilde{x}_\text{at},\widetilde{p}_\text{at})^T$,
$\mathbf{\widetilde{B}}_\gamma$ the covariance matrix for the photons,
$\mathbf{\widetilde{y}}_2 =
(\widetilde{x}_\text{ph},\widetilde{p}_\text{ph})^T$, and
$\mathbf{\widetilde{C}}_\gamma$ the correlation matrix for
$\mathbf{\widetilde{y}}_1$ and $\mathbf{\widetilde{y}}_2^T$. The
measurement of $\widetilde{x}_\text{ph}$ then transforms these
matrices according to
\begin{align}
  \label{eq:14}
  \mathbf{\widetilde{A}}_\gamma &\to \mathbf{\widetilde{A}}_\gamma
  - \mathbf{\widetilde{C}}_\gamma(\pi\mathbf{\widetilde{B}}_\gamma\pi)^-
  \mathbf{\widetilde{C}}_\gamma^T\\
  \label{eq:15}
  \mathbf{\widetilde{B}}_\gamma &\to \mathbbm{1}_{2\times2}\\
  \label{eq:16}
  \mathbf{\widetilde{C}}_\gamma &\to 0\\
  \label{eq:20}
  \mathbf{\widetilde{m}}_A &\to \mathbf{\widetilde{m}}_A
  + \mathbf{\widetilde{C}}_\gamma(\pi\mathbf{\widetilde{B}}_\gamma\pi)^-
  (x_\text{meas}-\widetilde{m}_4,0)^T\\
%  \mathbf{\widetilde{m}}_A &\to \mathbf{\widetilde{m}}_A
%  + \mathbf{\widetilde{C}}_\gamma(\pi\mathbf{\widetilde{B}}_\gamma\pi)^-
%  (\kappa \sqrt{\tau} m_2 + dW_2 - \kappa \sqrt{\tau} \widetilde{m}_3,0)^T\\
  \label{eq:21}
  \mathbf{\widetilde{m}}_B &\to 0
\end{align}

These equations fully describe the conditioned and the deterministic
evolution of the multi-variable Gaussian distribution, and in
particular we get access to the estimator for the $B$-field in the
form of its mean and the corresponding covariance matrix element. The
input to the estimation protocol is the constant parameters of the
problem and the outcome of the photo detection, which will drive the
mean values, and hence the estimator, to a non-trivial result,
cf.~Eq.~\eqref{eq:20}. We note that whereas the estimator depends on
the actual measurement outcome, the covariance matrix evolves in an
entirely deterministic manner, and we can hence theoretically predict
the magnitude of the error as it was also done
in~\cite{stockton04:_robus_quant_param_estim}.

We wish to simulate the protocol with a given realization of the noisy
field, and to this end we have to combine the theories of this and the
previous section. In such a simulation it is the actual field $B(t)$
that acts on the atoms and hence leads to the probability distribution
for the photo detection record, and the measured quantum field
variable $x_\text{ph}$ has the mean value $m_3=\kappa\sqrt{\tau}m_2$,
given by the expressions of the previous section. Hence the
measurement outcome can be written $\kappa\sqrt{\tau}m_2+\chi$ where
$\chi$ is uncorrelated with its value at previous detection times, and
it has vanishing mean and a variance of $1/2$. We are thus effectively
communicating the value of the simulated $B$-field through the use of
the ``theorist's'' estimator of the atomic state in the simulation of
measurement outcomes. Assuming a random measurement outcome with the
statistics described, we update the mean value vector and covariance
matrix of the previous subsection, and we use the same simulated
measurement outcome in the update formula~\eqref{eq:20}. The simulated
measurement outcome may deviate from the value $\widetilde{m}_4$
currently expected by the ``experimentalist'' both because of the
noisy contribution and because the mean is given by its true value and
not by his or her estimate. The latter is responsible for driving the
$B$-field estimate towards the actual realization in our numerical
simulations.

\section{Results}
\label{sec:results}

\begin{figure}
  \centering
  \includegraphics{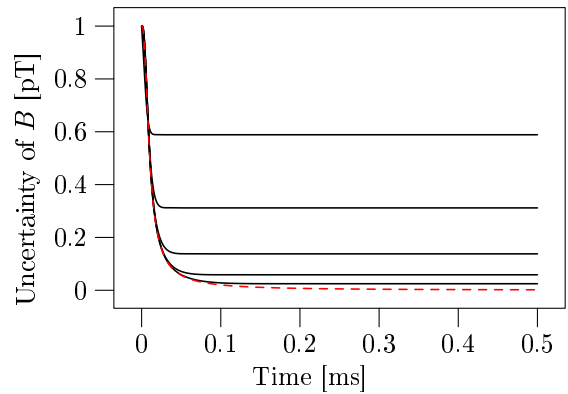}
  \caption{Uncertainty of the $B$-field as a function of time. We have
    used the numerical values of Table 1. The dashed/red line is the
    analytical result for a constant $B$-field. The full/black lines
    represents fluctuating fields with the same steady state
    uncertainty $\sqrt{\sigma_b/2\gamma_b}=1 \textrm{pT}$, but with
    decreasing rates. From above, $\gamma_b= \SI{10^5}{s^{-1}},
    \SI{10^4}{s^{-1}}, \SI{10^3}{s^{-1}}, \SI{10^2}{s^{-1}},
    \SI{10^1}{s^{-1}}$. For small values of $\gamma_B$ the full/black
    curves approach the constant $B$-field result as expected.}
  \label{fig:2}
\end{figure}

We now have a complete theory which, given the detection record,
provides the estimator for the time dependent $B$-field and its
variance. In practice, the experiment should be run, and the analysis
should be applied on the full detection record, either simultaneously
with the experiment or afterwards. Some calculation is necessary since
the $B$-field estimator at any time involves knowledge of the full
vector of mean values and the covariance matrix. The covariance
matrix, and in particular the variance of the $B$-field estimate
evolves deterministically with time. We can in fact solve the
equations for the covariance matrix analytically, and in the long time
limit we find the steady state variance on our $B$-field estimate:
\begin{gather}
  \label{eq:18}
  \begin{split}
    \var(B) &= \frac{1}{4\kappa^2\mu^2}
    \Bigl(\sqrt{\gamma_b^2+2\mu\sqrt{\sigma_{b}}\kappa}-\gamma_b\Bigr)^2\\
    &\relphantom{=} \times\sqrt{\gamma_b^2+2\mu\sqrt{\sigma_{b}}\kappa}.
  \end{split}
\end{gather}
When the Ornstein-Uhlenbeck process is slow, early detection events
and estimates provide already an estimate for future values of the
field, cf.~\eqref{eq:futureou}, which is further refined by the
continued measurement record. As also shown by~\eqref{eq:futureou}, if
the rate $\gamma_b$ is high, the estimates quickly loose their
significance, and only probing for a short time before $t$ is useful
in the estimate of of $B(t)$. In Fig.~\ref{fig:2} we show how the
variance of the $B$-field estimate starts with the prior steady state
value~\eqref{eq:steadyou} before any probing takes place, and evolves
to the steady state value given by~\eqref{eq:18}. In the figure we
assume the same steady state value for the $B$-field variance
$\sigma_b/2\gamma_b$ in all curves, but with different values of the
fluctuation rate constant $\gamma_b$. For comparison we show also in
the figure the analytical result~\eqref{eq:bvaranalyt} for a constant,
unknown field with the same prior uncertainty. This curve converges to
zero, but it is clearly seen to follow the curves with finite
$\gamma_b$ on short time scales (determined by $\gamma_b$).

\begin{figure}
  \centering
  \includegraphics{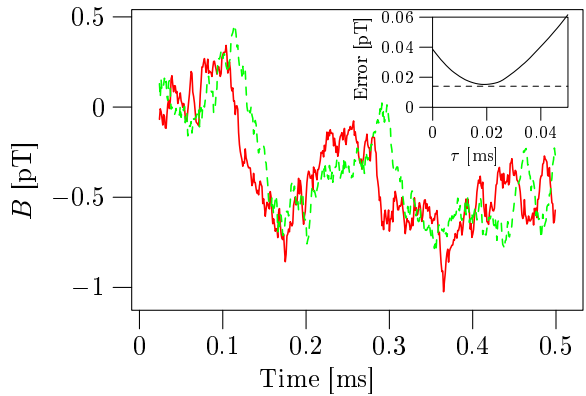}
  \caption{Time dependent value of the actual, simulated $B$-field
    (full/red line) and the estimate based on a simulated detection
    record (dashed/green line). The atomic parameters are given in
    Table 1, and the $B$-field is characterized by $\gamma_b =
    \SI{10^3}{s^{-1}} $ and $\sigma_b = \SI{2e3}{pT^2/s}$. The inset
    shows the error of the estimate~\eqref{eq:17} as a function of a
    variable translation of the two curves with respect to each other,
    and the horizontal dashed line in the inset shows the error
    obtained from a weighted average of delayed estimates, see text
    and Eq.~\eqref{eq:19}.}
  \label{fig:3}
\end{figure}

In Fig.~\ref{fig:3} we show the result of the application of a given
realization of the Ornstein-Uhlenbeck process. The figure shows the
actual process as a full line and the estimator as a dashed line for a
typical time window. We observe that the estimator tracks the gross
structure of the dependence very well, but smaller transients are not
reproduced. Taking a second look at the figure, we also observe that
the estimate is systematically lagging behind the true realization of
the field. This is not surprising, as the estimate makes explicit use
of all past measurements to predict the current value. In particular,
if no measurement results are provided, the estimate for the value is
simply obtained from past values according to Eq.~\eqref{eq:1}. This
should, however, prompt attempts to make an even better estimate for
the field using not only the detection record until the instant of
interest, but also the later detection events. We recall that we are
actually probing the atomic spin state, and the Larmor precession due
to the $B$-field may well be detected also after it took place.

To justify this increased effort, we have made a very simple
transformation of the data, consisting in a temporal displacement of
the two curves in Fig.~\ref{fig:3}, so that we compare the current estimator
$\widetilde{m}_1(t)$, obtained by our theory with earlier values of
the field $B(t-T)$. The inset of Fig.~\ref{fig:3} shows the numerically
calculated error of the estimate, averaged over a long detection
record, plotted as a function of the delay $T$,
\begin{gather}
  \label{eq:17}
  \mathrm{Error}^2(T) =  \frac{1}{t_0} \int_{T}^{t_0+T}
    (\widetilde{m}_1(t)-B(t-T))^2 dt,
\end{gather}
and we indeed see, that for a range of values for $T$ the error is
smaller than for $T=0$, i.e., we obtain a better estimate if we assign
our estimate to the value of the $B$-field a little earlier.

\begin{figure}
  \centering
  \includegraphics{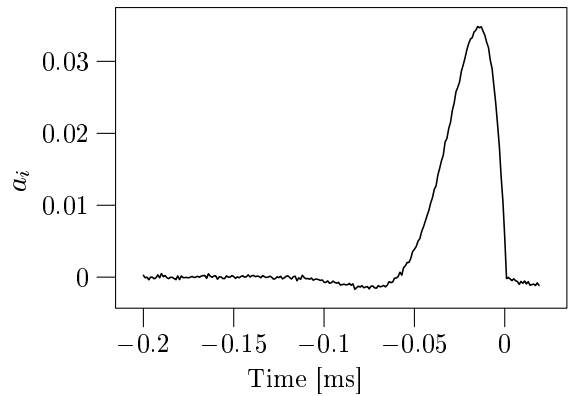}
  \caption{Time dependence of optimal weight factors $a_i$ in
    Eq.~\eqref{eq:19}, favoring contributions with a finite delay
    around 0.02 ms as suggested by the insert in Fig.~\ref{fig:3}.The
    atomic parameters are given in Table~\ref{tab:1}, and the
    $B$-field is characterized by $\gamma_b = \SI{10^3}{s^{-1}} $ and
    $\sigma_b = \SI{2e3}{pT^2/s}$.}
  \label{fig:4}
\end{figure}

A slightly more elaborate procedure to improve the estimate obtained
from the above procedure is obtained by assuming not just a simple
delay but a temporal convolution of the estimators,
$B_{opt}^{est}(t)=\int_{-\infty}^\infty a(t-t') \widetilde{m}_1(t')
dt'$. Numerically, we have identified the optimum delay distribution
by minimizing the (squared) error
\begin{gather}
  \label{eq:19}
 \frac{1}{t_0} \int_{T}^{t_0+T}
    \Bigl( \sum_i a_i \widetilde{m}_1(t-idt)-B(t) \Bigr)^2 dt,
\end{gather}
which turns out to be a linear algebra problem for the coefficients
$a_i$. Running a series of simulations, we find the temporal variation
of the delay distribution plotted in Fig.~\ref{fig:4}, peaking, as
expected, around the optimum delay, found in Fig.~\ref{fig:3}. The
error obtained this way is smaller than by use of any fixed delay, as
indicated by the horizontal dashed line in the inset in
Fig.~\ref{fig:3}.

\subsection{``Gaussian theory of hindsight''}
\label{sec:gauss-theory-hinds}

Both the fixed delay and the weighted average of delays bring promises
for improved sensitivity of magnetometers, if one can wait for the
estimate until (a short time) after the action of the field. Both
procedures, however, suffer from their ad hoc character, and in
particular from the fact that the optimum delay or delay distribution
are not theoretically available, unless one accepts to use simulations
as the present ones a guideline. We shall now present the correct
theory, which gives the optimum estimate and a tight bound on the
error without any ad hoc procedures.

The current estimate of the $B$-field is given by a Gaussian
probability distribution, and so is the estimate of the value at all
points in the past, and as measurements on the atoms proceed, we may
keep improving also our past estimate. We hence treat not only the
current value but also past values of the B-field as Gaussian
variables together with the atom and field variables. To update the
estimate at an instant $T$ in the past, we need to keep track of the
entire interval from $t-T$ to the present time $t$, and we extend our
Gaussian state formalism by replacing the argument $\widetilde{y}_1$
by a whole vector of values, representing the unknown $B$-field at
discrete times spanning an interval from $t-T$ until $t$. Since we are
dealing with the values in the past, they are not evolving due to the
Ornstein-Uhlenbeck process, but inherit their randomly value as time
proceeds and the elements in the vector of mean values and the
covariance matrix are simply ``pushed towards the past''. Only the
current value of $B(t)$ is given by the stochastic process. The atomic
system is correlated with both the current and the previous values of
the field as witnessed by non-vanishing elements in the extended
covariance matrix, and hence the updating due to measurements on the
atoms also influence the variance of the B-field estimate in the past,
c.f. the appropriate generalization of Eq.~\eqref{eq:20}. Now, there
is nothing ad hoc about the procedure. At any time, we have an
estimator for the value of the field over a finite interval looking
backwards in time, and we have the variance of these values, which is
a decreasing function of the time difference, approaching a constant,
for values so long time ago, that their action on the atoms is no
longer discernible, and hence no further updating takes place due to
Eq.~\eqref{eq:20}.

\begin{figure}
  \centering
  \includegraphics{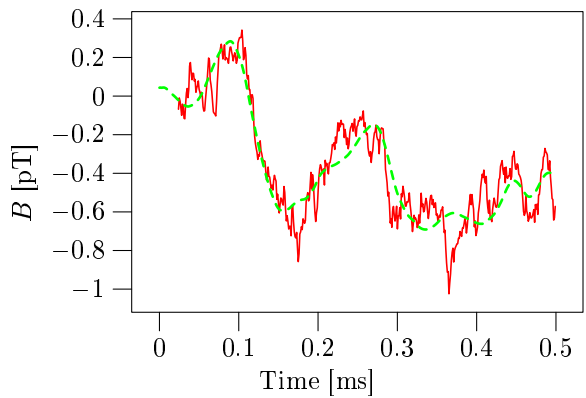}
  \caption{Time dependent value of the actual, simulated $B$-field
    (full/red line) and the estimate based on a simulated detection
    record available until 0.1 ms after the action of the field
    (dashed/green line). The atomic parameters are given in Table 1,
    and the $B$-field is characterized by $\gamma_b =
    \SI{10^3}{s^{-1}} $ and $\sigma_b = \SI{2e3}{pT^2/s}$.}
  \label{fig:5}
\end{figure}

Fig.~\ref{fig:5} shows a comparison of the actual, simulated field
with our Gaussian estimator, available 0.1 ms after the action of the
$B$-field on the atoms. Rapid transients are not reproduced, but in
comparison with Fig. 3, we have clearly removed the lag and improved
the overall agreement between the estimator and the actual value of
the time dependent field.

\begin{figure}
  \centering
  \includegraphics{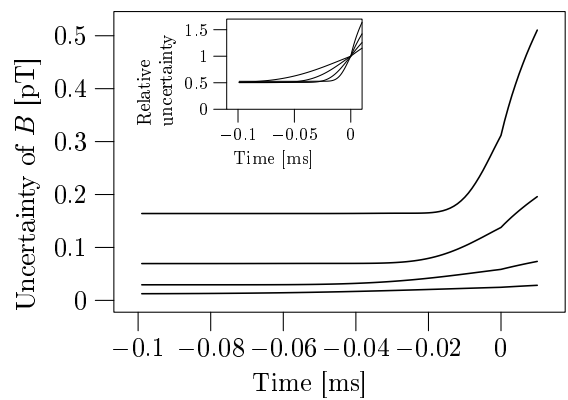}
  \caption{Time dependence of the variance of our estimate of the time
    dependent field $B(t)$-field around the current time $t_L$ in the
    laboratory. The curves represent fluctuating fields with the same
    steady state uncertainty $\sqrt{\sigma_b/2\gamma_b}=1
    \textrm{pT}$, but with decreasing rates. From above, $\gamma_b=
    \SI{10^4}{s^{-1}}, \SI{10^3}{s^{-1}}, \SI{10^2}{s^{-1}},
    \SI{10^1}{s^{-1}}$. The atomic parameters are given in Table 1.
    For later times $t>t_L$, we have to guess the value from the
    properties of the Ornstein-Uhlenbeck process~\eqref{eq:futureou},
    at the current time we have the value~\eqref{eq:18}, and for
    earlier times we get the improvement given by the solution of the
    extended Gaussian state updating. The inset shows the uncertainty
    relative to its value at $t=t_L$; improved by an approximate
    factor of 2, when we estimate the past values of the field.}
  \label{fig:6}
\end{figure}

Fig.~\ref{fig:6} reports the variance of the estimate of the
$B(t)$-field plotted at a given time $t_L$ in the laboratory after a
long measurement time, so transients in the atomic dynamics have died
out. Results are shown for different values of $\gamma_b$ but a
constant ratio between $\gamma_b$ and $\sigma_b$. The estimate for
$B(t=t_L)$ at the laboratory time is based on measurements that have
registered the action of the field at all previous times and takes the
value given by~\eqref{eq:18}. For future times $t > t_L$, we have to
use the current estimate and extend it by our knowledge of the
Ornstein-Uhlenbeck process, i.e. by inserting $V_{t_L}$ in
Eq.~\eqref{eq:futureou}. The curves show that it is easier to guess
the present value than future ones, and that future values are
particularly hard to guess for a rapidly fluctuating process. Finally,
the curves also explore the range of $t < t_L$, where we estimate past
values of the field, and we observe the value of hindsight. For small
$\gamma_b$, the uncertainty is small, but we see in the inset of the
figure, where the variance is plotted relative to its value at
$t=t_L$, that quite independently of the time constant of the
fluctuations, if we can only wait long enough, we find an improvement
by a factor of approximately 2 on the uncertainty compared to the
equal time estimate. This improvement is a function of the physical
coupling parameters, and the present study suggests a careful analysis
of the optimum probing strategy, including the possibility of time
dependent probing field strengths and detunings and taking into
account also atomic decay processes.

\section{Discussion}
\label{sec:discussion}

We have presented a formalism, that provides the correct estimate of a
time dependent $B$ field which is known to fluctuate randomly
according to an Ornstein-Uhlenbeck process. The estimate exhausts the
measurement data and makes the tightest possible conclusions from the
photo detection record. It is ``correct'' in the sense that any better
estimate necessarily requires further information, which could either
be in the form of prior information about the field or the results of
further measurements on the system. Our ability at time $t$ to
estimate $B(t)$ connects in a natural manner with previous results for
the estimation of static fields, and it is comparable with the results
of the analysis of time dependent fields
in~\cite{stockton04:_robus_quant_param_estim}. Simulations show that
this theory actually provides a better estimate of the value in the
recent past (or, instrumentally, if you need to estimate the field at
time $t$, you should use the formally obtained estimate at time $t+T$
for some suitable $T$). Our formalism naturally generalizes to
describe also a scenario, where previous estimates are updated by
current measurements, and this provides an essential improvement for
magnetometers as documented in the paper.

We have not made a complete survey of the optimal performance as a
function of all physical parameters. At this stage, it seems futile to
vary all parameters without inclusion, in particular, of atomic decay,
which will either restrict the magnetometers to finite time analyses
or which will have to be compensated by a continuous optical
re-pumping of the atoms. As pointed out
in~\cite{stockton04:_robus_quant_param_estim}, the field estimation
can be made robust to imprecise information about some of the physical
parameters, in particular the number of atoms which enters the
atom-light interaction strength. Instead of just calculating the
field, one can apply, in a feed back set-up, a compensating field
which should freeze the Larmor precession, independently of the number
of atoms, and the field estimate is now given by the value of this
compensating field. It is possible to include a time dependent feed
back field in our equations, and hence to simulate the performance of
any feed back strategy and its robustness against fluctuations in
physical parameters. Another important extension of the theory deals
with the assumption of the Ornstein-Uhlenbeck process with given
parameters. The parameters enter explicitly in our estimation
procedure, and if these parameters are not known, or if they are
functions of time, e.g., as in the case of cardiography on a beating
heart, more refined theory will bee needed. At this point, we recall,
that the Ornstein-Uhlenbeck model represents our prior knowledge about
the field fluctuations. A conservative high estimate on the noise
fluctuations will hence give a similar conservative estimate on the
field, which might be improved if better limits were known. A
theoretical investigation of this problem could for example simulate
the estimation of a noisy field, assuming a different value of the
parameters than applied in the synthesis of the field, and investigate
numerically the difference between the variance on the estimate
obtained from the theory and the actual statistical agreement between
the estimate and the field.

\begin{acknowledgments}
  The authors are grateful to the ONR-MURI collaboration on quantum
  metrology with atomic systems and to L.~B.~Madsen for fruitful
  discussions in the early stages of this project.
\end{acknowledgments}

% Create the reference section using BibTeX:
%\bibliography{bibliography}

\end{document}